\documentclass[%
 reprint,
 amsmath,amssymb,
 aps,
]{revtex4-2}

\usepackage{graphicx}
\usepackage{dcolumn}
\usepackage{bm}
\usepackage{xcolor}

\begin{document}

\preprint{APS/123-QED}

\title{Surface-State Dissipation in Confined $^3$He-A}

\author{Alexander J. Shook}
\email{ashook@ualberta.ca}
\author{Emil Varga}%
\author{Igor Boettcher}
\author{John P. Davis}%
\email{jdavis@ualberta.ca}
\affiliation{%
Department of Physics, University of Alberta, Edmonton, Alberta, Canada T6G 2E9
}%


\begin{abstract}
We have studied the power dependence of superfluid Helmholtz resonators in flat (750 and 1800 nm) rectangular channels. In the A-phase of superfluid $^3$He, we observe a non-linear response for velocities larger than a critical value. The small size of the channels stabilizes a static uniform texture that eliminates dissipative processes produced by changes in the texture. For such a static texture, the lowest velocity dissipative process is due to the pumping of surface bound states into the bulk liquid. We show that the temperature dependence of the critical velocity observed in our devices is consistent with this surface-state dissipation. Characterization of the force-velocity curves of our devices may provide a platform for studying the physics of exotic surface bound states in superfluid $^3$He.
\end{abstract}

\maketitle

One of the defining features of superfluidity is the ability to flow without dissipation for velocities below a critical value \cite{khalatnikov2018introduction}. The Landau criterion states that this velocity threshold is set by a local minimum in the dispersion relation of the lowest energy excitation of the system \cite{khalatnikov2018introduction}. For fermionic superfluids, the relevant energy scale is the superfluid gap, $\Delta_{\vec{p}}$, which is the energy required to excite a quasiparticle from the Fermi surface, and the Landau critical velocity is therefore $v_L = \Delta_{\vec{p}}/p_F$ \cite{Vollhardt2013}.

Implicit in the arguments of Landau is the assumption that the gap is both spatially homogeneous and isotropic. In superfluid $^3$He the latter assumption holds only for the bulk B-phase, which has an isotropic gap. Near a surface, however, the gap is suppressed and develops separate parallel and perpendicular components \cite{Vollhardt2013,rudd2021strong}. The suppression of the gap near the wall breaks the Landau assumption and allows for bound states with energies less than the bulk gap. Experiments studying oscillating macroscopic objects in $^3$He-B have shown that there is a sub-Landau critical velocity threshold at which bound states are emitted from a moving surface, leading to an observable change in dissipation \cite{castelijns1986landau,bradley2016breaking}. Characterization of the coupling of these mechanical oscillators to fluid flow has proven to be a valuable tool for studying surface bound states in $^3$He-B \cite{castelijns1986landau,skyba2011high,zheng2017critical,bradley2016breaking,noble2022producing,autti2020fundamental,autti2023quasiparticle}, which supplements other techniques \cite{autti2020exceeding,lotnyk2020thermal,scott2023magnetic}. These surface bound states are of interest not only from the perspective of understanding $^3$He hydrodynamics \cite{kuorelahti2018models} and quantum turbulence \cite{fisher2014andreev,tsepelin2017visualization}, but also as a condensed matter realization of exotic quasiparticles such as Weyl or Majorana fermions \cite{chung2009detecting,tsutsumi2010majorana,silaev2012topological,ikegami2013chiral,park2015surface,shevtsov2016electron,volovik2017chiral,vorontsov2018andreev,byun2018measuring,wu2023weyl}. Experimental studies of surface bound state dissipation have thus far been limited to the B-phase, as research on A-phase surface-states is more complex due to the intrinsic anisotropy of the gap. Here, we reveal the pumping of surface bound states into the bulk A-phase, with a lower critical velocity than the B-phase due to  additional suppressed states from the anisotropic A-phase gap.

\begin{figure}[b]
\includegraphics[width=\linewidth]{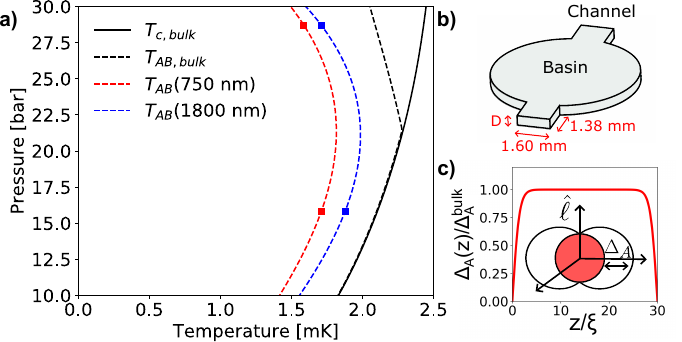}
\caption{\label{fig:Fig1} A-phase flow. (a) Simplified phase diagram showing the range of temperatures where the A-phase exists for 750 and 1800 nm channels. The red and blue points are the transition temperatures measured for both devices and the dashed lines are fits. The A-phase is stabilized to lower temperatures and pressures by the tight confinement \cite{shook2020stabilized}. (b) Drawing of the confined Helmholtz resonator volume. The dimensions of the channel are $D$ $\times$1.6 mm $\times$ 1.38 mm, where $D=$ 750 or 1800 nm. (c) Momentum space plot of the Fermi surface (red) and the gap which goes to zero at the poles aligned with the anisotropy vector $\hat{\ell}$. The plot shows the spatial dependence of $\Delta_A(z)$, in the highly confined dimension, where the distance from the wall is in units of the coherence length $\xi$. The spatial dependence has been computed using Ginzburg Landau theory \cite{Vollhardt2013}.}
\end{figure}

In the A-phase, there exists an anisotropy axis $\hat{\ell}$, which points in the direction of a Cooper pair's orbital angular momentum, along which the superfluid gap closes at two point nodes (see Fig.~\ref{fig:Fig1}). Since the A-phase exhibits long-range spatial ordering in the orbital angular momentum of Cooper pairs, $\hat{\ell}(\vec{r})$ is a vector field, or texture, which can vary smoothly in space over distances larger than the coherence length. The magnitude of the gap for a given momentum, $\vec{p}$, is specified by the equation $\Delta_{\vec{p}} = \Delta_A [1-(\hat{p}\cdot\hat{\ell})^2]^{1/2}= \Delta_A |\sin\theta|$. Here, $\theta$ is the angle between the momentum and anisotropy axis, such that the gap has a maximum magnitude, $\Delta_A$, for excitations with momentum $\vec{p}\perp\hat{\ell}$ and a minimum magnitude of zero for $\vec{p}\parallel \hat{\ell}$. For this reason, naive application of the Landau criterion implies a critical velocity of zero.

Critical velocities of a different kind are possible in cases where the texture is dynamic. The motion of $\hat{\ell}$ changes the gap, and therefore dissipates energy by creating new excitations as it moves \cite{cross1977orbital}. The texture couples both to superfluid phase gradients (i.e.,~flow), and to spin degrees of freedom \cite{Vollhardt2013}. In the absence of other orientational effects, the tendency of the $\hat{\ell}$-texture is to align with the superfluid flow velocity $\vec{v}_s$ \cite{de1974alignment}. This tendency is in competition with the boundary conditions, which requires the $\hat{\ell}$-texture to be perpendicular to surfaces. This means that $^3$He-A flowing over a surface can produce a textural gradient where $\hat{\ell}$ is parallel to the flow far from the wall and perpendicular at the surface \cite{de1974alignment}. The characteristic length scale over which the texture rotates by $90$ degrees is the healing length, $\xi^A_{\textrm{heal}} \sim 8$ $\mu$m \cite{Vollhardt2013}. For bulk systems, where all dimensions are large compared to the healing length, the texture becomes a hydrodynamic variable that exhibits complicated behavior, including critical velocities \cite{volovik1976hydrodynamics,mermin1976circulation,bhattacharyya1977stability,cross1978stability, fetter1978hydrodynamic,kleinert1979two}. Systems where one or more dimensions are small compared to the textural healing length tend to lock a particular texture in place. This can be seen in the literature from experiments with varying degrees of confinement \cite{johnson1975heat,bagley1978direct,parpia1979critical,dahm1980dissipation,paalanen1980dissipation,manninen1982critical,kotsubo1987suppression,daunt1988critical,kasai2018chiral,manninen1983flow,thuneberg1981difficulties}.
 
In cases where the texture is static, constant A-phase superfluid flow can be stable even when aligned with $\hat{\ell}$, because only a small number of states exist near the nodes. These states quickly fill when the fluid begins to flow, but once filled do not contribute to dissipation \cite{volovik1984superfluid}. This produces a non-linear relationship between the superfluid velocity and momentum density,
\begin{equation}
    \vec{j}_s = \rho_s(v_s) \vec{v}_s.
\end{equation}
The superfluid density, $\rho_s$, decreases with increasing velocity as excitations are produced. Thus the momentum density, $j_s(v_s)$, has a local maximum known as the maximum pair breaking current \cite{Vollhardt2013}. This relationship assumes the system is always near equilibrium such that the available states are  filled. There is, therefore, no special velocity at which dissipation onsets for this static texture, constant flow, case.

Until now, an open question remained as to what critical velocities, if any, the A-phase would exhibit if the texture is stationary but the flow is oscillatory. As with the DC flow case, excitations can be produced near the nodes for arbitrarily small velocities, but to continuously dissipate energy there must be some process by which these states are continuously populated, and then emptied. The most obvious candidate is the bound state pumping process already known to exist in the B-phase \cite{castelijns1986landau,bradley2016breaking}.

Our experiment studies the critical dissipative behavior of the A-phase using alternating flow in a parallel plate geometry with confinement much smaller than the healing length \cite{zhelev2017ab,levitin2019evidence,shook2020stabilized,sun2023superfluid,yapa2022triangular}, ensuring a uniform texture. We have made use of our nanofluidic devices called Helmholtz resonators, which have been described in previous publications \cite{rojas2015superfluid,varga2020observation,shook2020stabilized,varga2021electromechanical}. The devices are comprised of bonded quartz chips that have been etched to create a small volume sandwiched between the chips. The shape of this space is a circular basin (3.5 mm radius), with two 1.60$\times$1.38 mm$\times D$ rectangular channels connecting it to the external helium bath. The variable $D$ is the thickness of the enclosed space, which is constant throughout. The two devices used in this experiment had thicknesses of $D =$ 750 $\pm$ 12 and 1800 $\pm$ 12 nm.

Aluminum electrodes are patterned onto the quartz, creating a parallel plate capacitor inside the basin. The volume of the basin can be slightly decreased by an electrostatic force between the capacitor plates. When this plate motion is driven resonantly with the Helmholtz mode of the channels, fourth sound is driven. The normal fluid does not move in the channels because the viscous penetration depth, $\delta$, is large compared to the confinement ($\delta\approx$ $400$ $\mu$m $\gg D$) \cite{reppy1977superfluid}. Furthermore, the confinement is also small compared to the healing length ($D \ll \xi^A_{\textrm{heal}}\approx 8\,\mu$m ) \cite{Vollhardt2013}. Therefore, the texture is uniformly aligned in the highly confined direction, $\hat{z}$, effectively eliminating textural dissipation mechanisms. 

\begin{figure}[b]
\includegraphics[width=\linewidth]{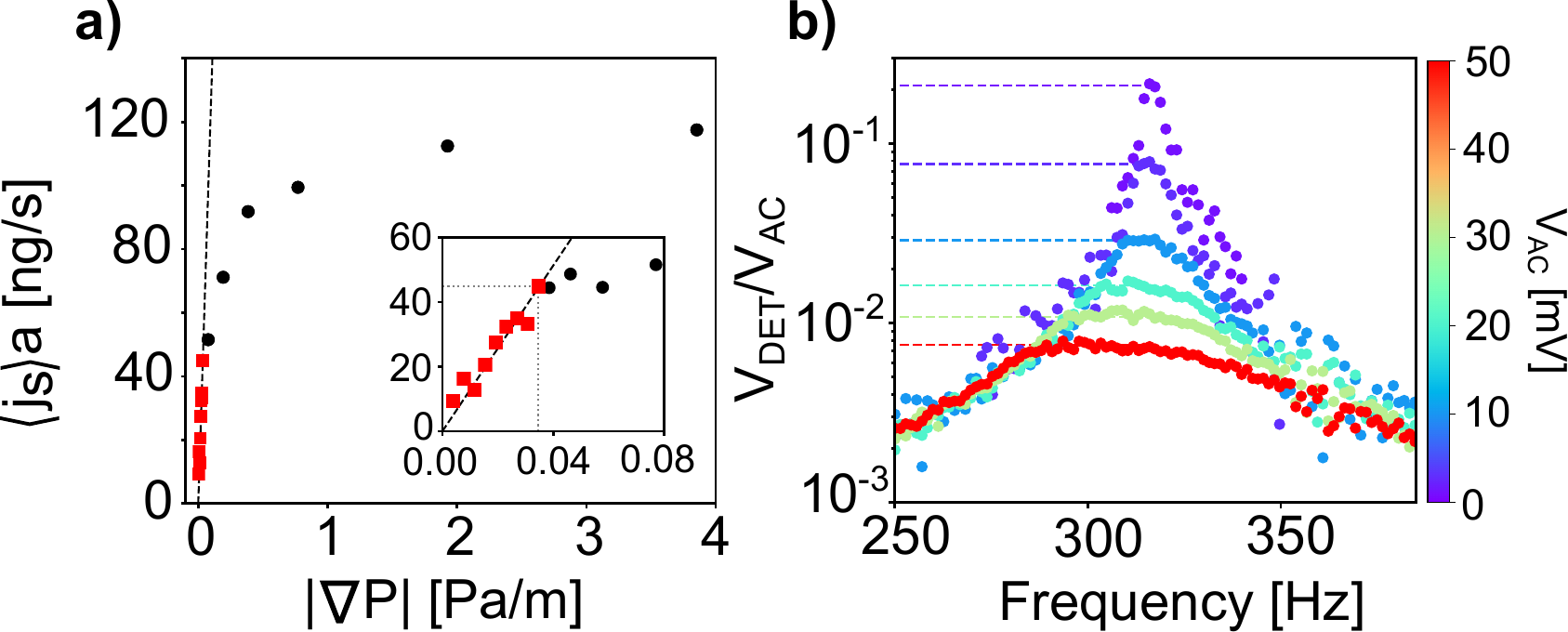}
\caption{\label{fig:Fig2} Characterizing the non-linear regime. (a) The mass current, calibrated from the detector peak voltage, for the 750 nm device at 22.45 bar and 2.08 mK is plotted as a function of the pressure gradient across the channels $|\nabla P|$. The bias voltage $V_{\textrm{DC}}$ is held constant throughout the experiment such that $\langle j_s \rangle a \propto V_{\textrm{DET}}$ and $|\nabla P| \propto V_{\textrm{AC}}$. The inset highlights the linear drive regime and a critical value at which the slope abruptly changes. The dashed line is a fit to the linear regime data used to highlight the change in slope. The dotted lines indicate the point at which the slope changes. (b) Log-plot of the Helmholtz resonance with drive voltages ranging from 1 to 50 mV. The resonances are normalized by the drive voltage. Near resonance, the line shape distorts above the critical drive.}
\end{figure}
The capacitance of the Helmholtz resonator varies in time when driven. The measured capacitance signal responds to changes in the basin fluid mass when the fourth sound resonance is driven. A model of this system, described in the Supplementary Material ~\cite{[{See Supplemental Material, which includes Refs. \cite{ichikawa1987healing,freeman1990size,landau2013fluid,vorontsov2003thermodynamic,buchanan1986velocity,bartkowiak2002unique,freedericksz1933forces,zurek1985cosmological,ruutu1997intrinsic,schwarz1990phase,barenghi2001quantized,bradley2008grid,walmsley2004intrinsic}, at }]supp}, relates the spatially averaged mass current, $\langle j_s \rangle$, to the measured detector voltage via the equation
\begin{equation}
    \langle j_s \rangle  = \frac{(1+2\Sigma)}{2C_0R_{\textrm{trans}}} \left( \frac{\rho A D}{a} \right) \frac{V_{\textrm{DET}}}{V_{\textrm{DC}}}.
    \label{eq:js}
\end{equation}
Here, $C_0$ is the undriven capacitance, $R_{\textrm{trans}}$ is the current to voltage conversion factor of a transimpedance amplifier, $\rho$ is the total mass density of the $^3$He, $V_{\textrm{DC}}$ is a bias voltage used to enhance the signal, $A$ is the area of the basin, $a$ is the cross-sectional area of the channel, and $\Sigma$ is a small correction factor to account for the finite compressibility of the helium.

By performing repeated power-sweep measurements of the Helmholtz resonance, we can measure the drive dependence of the resonance amplitude \cite{varga2020observation}. As shown in Fig.~2, for low drives there is a linear regime where the peak amplitude is proportional to the drive voltage, suggesting that the superfluid density is independent of drive. Once the peak of the resonances crosses a critical value, $V_c$, there is a secondary regime where the amplitude increases at a slower rate and the line shape begins to flatten at the top of the resonance. The amplitude saturates for large drive voltages, suggesting that there is a maximum momentum density beyond which we cannot drive the resonator.

\begin{figure}
\includegraphics[width=\linewidth]{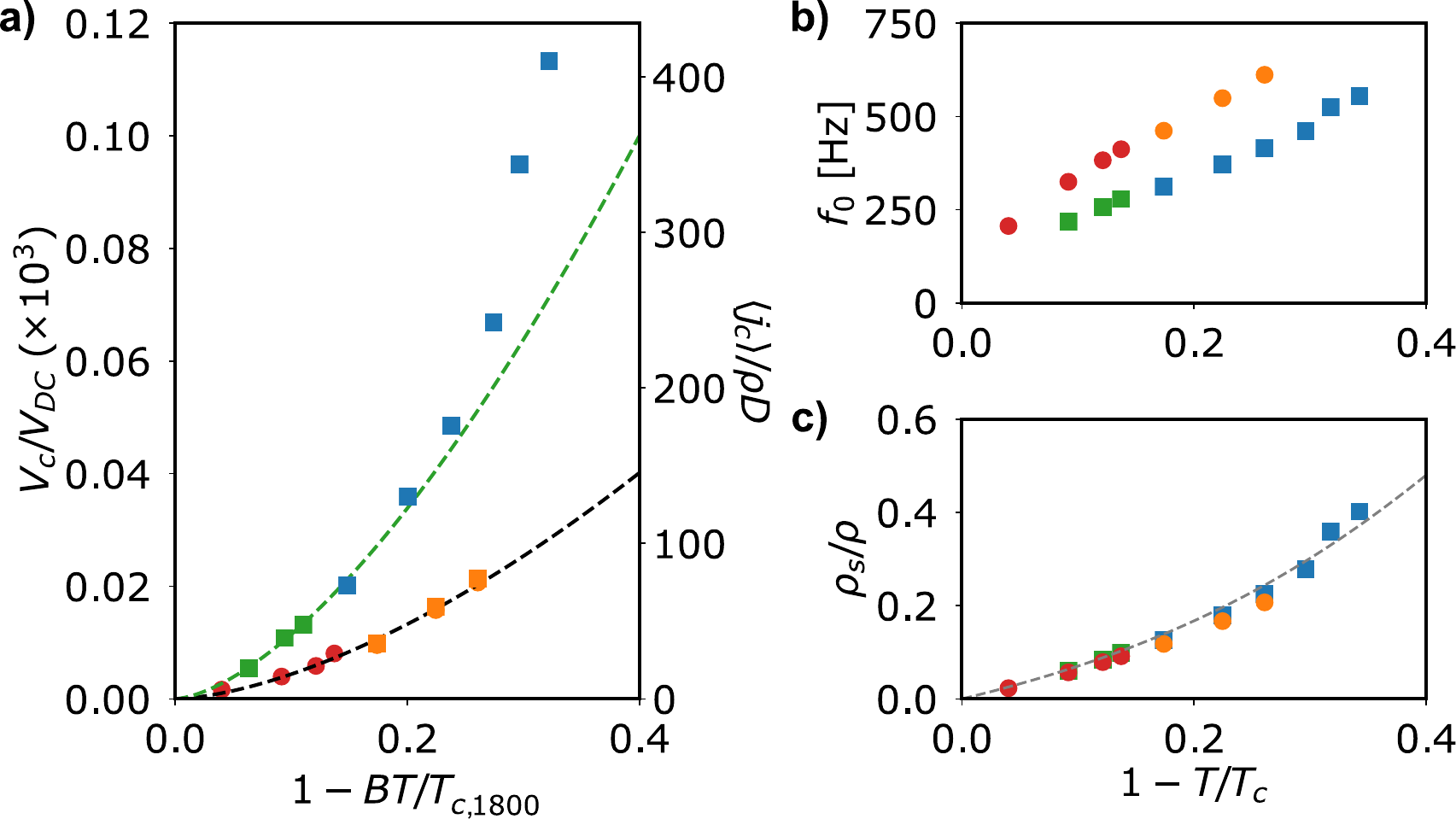}
\caption{Temperature scaling. (a) Plot of the peak voltage as a function of temperature for the 750 nm device at 22.45 bar (green squares), and 27.94 bar (blue squares), as well as 1800 nm device at 22.45 bar (red circles), and 27.94 bar (orange circles). The dashed lines show fits to functions of the form $(1-BT/T_{c,1800})^{n/2}$. The parameter $B$ is used to rescale the critical temperature for the 750 nm device. The 22.45 bar and 27.94 bar data sets are fit together for the 1800 nm device but separately for the 750 nm device. (b) Measured frequency dependence of the Helmholtz modes. (c) Superfluid density of each device as calculated from the resonant frequency. The grey curve is the bulk superfluid fraction.}
\label{fig:Fig3}
\end{figure}

We characterize this effect by recording the detector voltage at which the slope changes for both devices at pressures of 2.87, 22.45 and 27.94 bar, over a range of temperatures. This voltage threshold was then converted into a critical current using Eqn.~\ref{eq:js}. These results are compiled in Fig.~\ref{fig:Fig3}. The temperature was determined using the known temperature dependence of the superfluid density \cite{parpia1985temperature}, calibrated to a primary melting curve thermometer \cite{greywall19823}.  Specifically, the temperature of each data point was computed using the resonance frequency of the fourth sound mode of the 1800 nm device. The fourth sound mode frequency changes according to the equation
\begin{equation}
    \left( \frac{\omega_0(T)}{\omega_{0}(0)} \right)^2 = \frac{\rho_s}{\rho},
\end{equation}
where $\omega_0(0)$ is a function of the resonator dimensions, total fluid density, and the isothermal compressibility. Inversion of this curve allows the measured frequency to be converted into a temperature as described in the Supplementary Material~\cite{supp,rojas2015superfluid}. 

The temperature scaling was investigated by fitting 750 nm and 1800 nm amplitude data sets to a function of the form 
\begin{equation}
\langle j_c \rangle = j_0(1-BT/T_{c,1800})^{n/2}. 
\end{equation}
The prefactor $B$ is included to account for the suppression of the critical temperature due to confinement. For the 1800 nm device $B=1$, and for the 750 nm device it is the ratio of the two critical temperatures $B = T_{c,1800}/T_{c,750}=1.042$. The value of this ratio is inferred by measuring the mode frequency as a function of temperature and extrapolating to zero frequency.

\begin{figure}[b]
\includegraphics[width=\linewidth]{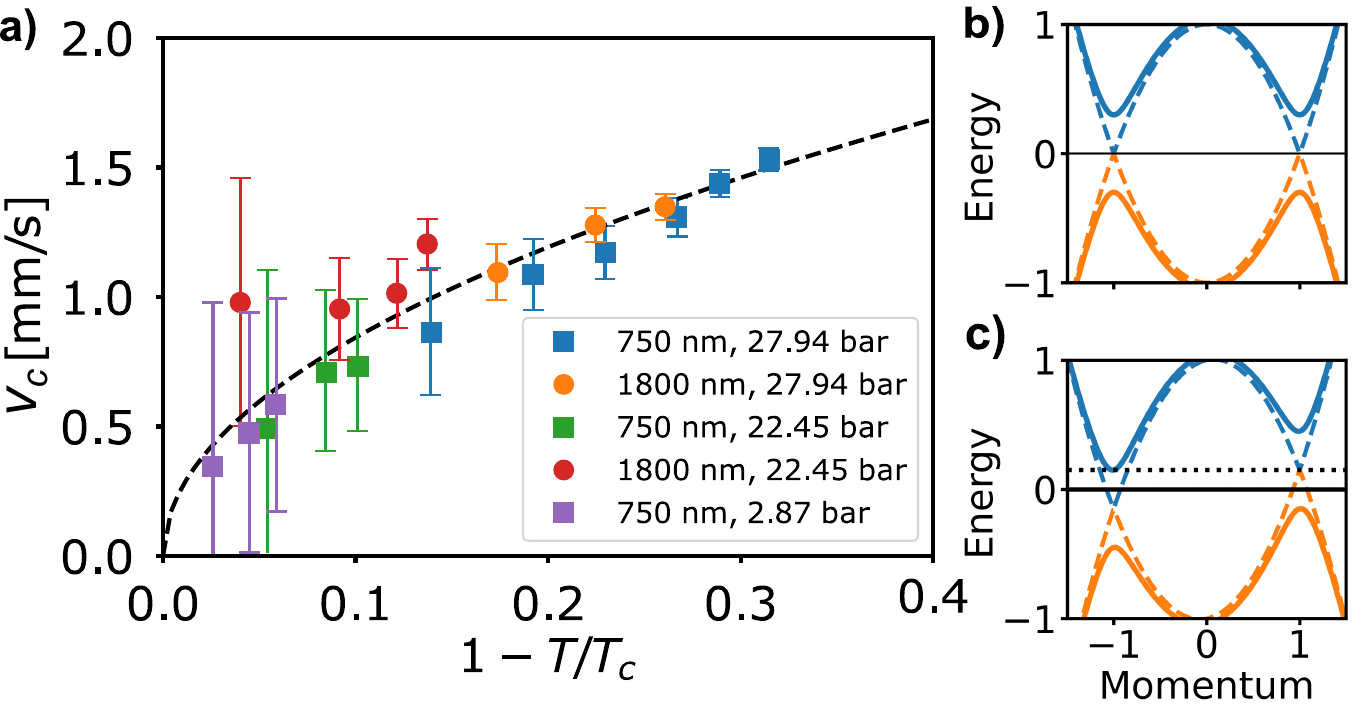}
\caption{Critical velocity. (a) The critical current of the Helmholtz resonator is computed from the resonance amplitude at which the linear regime ends, and the superfluid density from the center frequency of the resonance as $v_c = \langle j_s\rangle/\rho_{s,c}$. The ratio of these values is the critical velocity. The dashed line is a fit to all data sets of the form $v_{0}\sqrt{1-T/T_c}$. (b) Plot of the quasiparticle dispersion relation when the fluid is at rest. The solid lines represent the bulk dispersion relation, and the dashed lines the bound state dispersion. (c) Plot of the quasiparticle dispersion relation at a finite critical velocity when the maximum energy value of the lower-band bound states is equal to the minimum energy value of the upper-band bulk states (highlighted by the dotted line).}
\label{fig:Fig4}
\end{figure}

For the 1800 nm device, both pressure data sets are well fit by $n=3.20$. For the 750 nm device, a similar curve, $n=3.13$, can be fit to the data, but it deviates from this trend at lower temperatures. The fact that the A-phase persists to lower temperatures under higher confinement allows us to measure a wider range of temperatures in the 750 nm device. The deviation from a $3/2$ power law appears to be a consequence of the fact that the superfluid fraction is approximately linear near $T_c$, but not at lower temperatures.

The critical currents are converted into critical velocities, by taking the ratio $v_c = \langle j_c \rangle/\rho_{s,c}$. Here, $\rho_{s,c}=\rho_{s}(v_c)$ is the velocity-dependent superfluid density computed from the measured Helmholtz frequency at the critical value. We find the critical velocity curves for the two devices, at three different pressures, come close to falling onto one another (see Fig.~4) and are not inconsistent with a function of the form
\begin{equation}
    v_c = v_0(1-BT/T_{c,1800})^{1/2},
\end{equation}
which is the same temperature scaling as the Ginzburg-Landau gap. The temperature-independent prefactor of the fit is $v_0=2.65 \pm 0.09$ mm/s. This is reminiscent of the Landau critical velocity for an isotropic superfluid $v_L(T) = \Delta(T)/p_F$. However, the analogy is not straightforward due to the existence of the A-phase nodes.

Comparing our results to the DC flow experiments performed by Manninen et al.~\cite{manninen1982critical,manninen1983flow}, which studied flow through an $0.8$ $\mu$m Nuclepore filter, multiple dissipation regimes were observed only in cases where the end effects produced orbital viscosity. These end effects occurred only when the A-phase existed both inside and outside the pores, but not when the superfluid was B-phase outside the pores and A-phase inside. In light of this, it is worth considering the phase transitions of the bulk fluid outside the Helmholtz resonator. At 22.45 bar the bulk A to B transition occurs at $T_{AB} = 0.979 T_c$, red{at} 27.94 bar it is $T_{AB} = 0.876T_c$, and at 2.87 bar it does not occur at all. This means that in the majority of our measurements the fluid outside the Helmholtz resonator is B-phase. To study the role of the boundary, we performed a measurement at 27.94 bar at 2.37 mK, which is above the bulk $T_{\textrm{AB}}$ line, ensuring A-phase both inside and outside the Helmholtz resonator. We found the critical velocity follows the same temperature scaling, suggesting the phase boundary plays no role \cite{supp}. In the experiments of Ref.~\cite{manninen1982critical,manninen1983flow} where the A-phase texture was static, there was no special velocity at which dissipation onsets, as expected for DC flow. This suggests that the dissipation onset velocity we observe is unique to the dynamics of oscillatory flow resulting from our AC Helmholtz resonance. 

To understand the role of oscillatory flow, we now consider in more detail the mechanical oscillator experiments performed in $^3$He-B \cite{castelijns1986landau,bradley2016breaking}. Near the surface of a moving object, the gap is suppressed allowing for bound state excitations localized near the surface. Similar to the states near the A-phase nodes, these states do not contribute to dissipation once filled, unless they can escape into the bulk fluid. When the flow is alternating, with an oscillation period that is large compared to the quasiparticle lifetime, a pumping process can occur at a fraction of the Landau critical velocity when the energy of a bound state exceeds that of an unoccupied bulk state. This allows for dissipation as the surface-states are continuously populated and released. There does not seem to be any conceptual reason why this same process should not occur in the A-phase and, when textural dynamics are eliminated, we argue it should be the lowest energy critical velocity. Since the critical velocity due to bound state dissipation is proportional to the gap magnitude, this is consistent with the temperature scaling we observe. It is worth mentioning that although the previously mentioned mechanical oscillator experiments \cite{castelijns1986landau,bradley2016breaking} needed to cool to the ballistic temperature regime to study bound state dissipation, this is due primarily to the normal fluid viscosity. Since our experiment is based on a superleak, a critical velocity due to bound states could in principle be measured at any temperature, though the existence of thermal excitations may modify the results quantitatively (see discussion in Supplementary Material \cite{supp}).

We note that the flow inside the Helmholtz resonator is quite different from a vibrating wire. Due to viscous clamping of the normal fluid, there is no analogous backflow parameter for the Helmholtz resonator. Analysis of the flow fields, however, shows that there is localized flow enhancement at the corners of the channels. Our simulations (discussed in the Supplementary Material \cite{supp}) suggest that the peak flow velocity at the corners may be a factor of $\sim$10 times larger than the spatially averaged channel velocity that we calculate from our measurements. For this reason, bound state dissipation is likely initially localized to a region near the corners. Our experiment is only sensitive to the average velocity, therefore the values we report in Fig.~4a may not reflect the velocity at pair breaking \cite{lambert1992theory}. Future experiments will investigate the effects of rounded corners on the critical velocity. Such rounded corners will ensure uniform velocity, hence quantitative measurements of the pair breaking velocity.

\begin{figure}[t]
    \centering
    \includegraphics[width=0.9\linewidth]{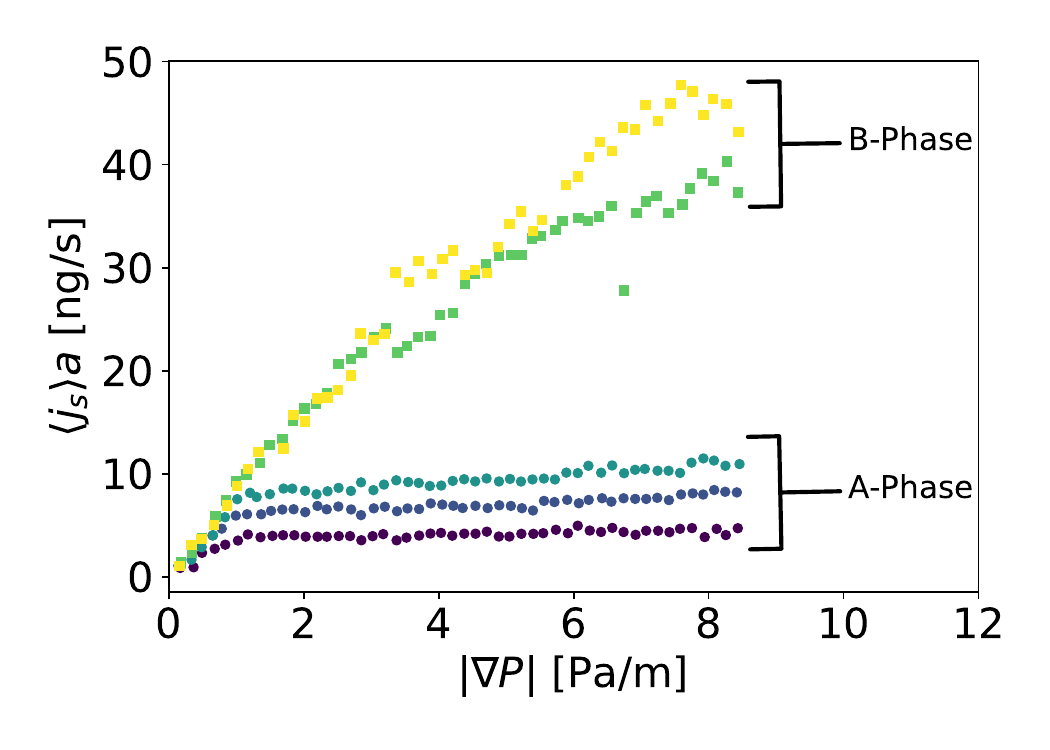}
    \caption{Plot of the resonance peak height as a function of drive voltage, measured for the 1800 nm device at 28 bar for a variety of temperatures near the A-B critical temperature. The A-phase data is plotted in circles, and the B-phase in squares. Lower temperature measurements have higher amplitudes due to the increased superfluid fraction. For low drives, the drive-amplitude relationship is linear for both A and B phases. Above the A-phase critical velocity, the trend deviates for the A-phase as the line shape becomes distorted with a flat top, whereas in the B-phase this flattening does not occur.}
    \label{fig:enter-label}
\end{figure}

We argue in the Supplementary Material that the confinement does not permit textural transitions even when the flow enhancement is considered \cite{supp}. The question of dissipation due to vortices (i.e.~nucleation at the channel corners, or a vortex mill process) is also explored in the Supplementary Material \cite{supp}, which has been considered as an alternative interpretation of the observed critical velocity. We offer several arguments against this view, the most important of which is the difference in the non-linear scaling we observe in the B-phase compared to the A-phase. Any dissipative process involving pure phase winding vortices, which exist in both the A and B-phases, should give rise to a similar critical velocity in both phases. Contrary to this, we find a drastically different drive scaling for the B-phase. Figure 5 shows how the peak height of the Helmholtz resonance scales with the drive voltage for the 1800 nm device at 27.94 bar, for temperatures above and below the A-B transition temperature. In the A-phase there is an obvious critical drive where the slope changes as the resonance line shape begins to distort. For the B-phase the scaling remains linear for higher drives, eventually displaying a qualitatively different type of non-linearity. Instead of the resonance line shape becoming flatter, it develops a bistability, which will be explored elsewhere.

The fact that the B-phase does not display the same critical velocity as the A-phase suggests that the dissipation mechanism must either be unique to the A-phase or simply occur at higher velocities in the B-phase. A simple model has been constructed in the Supplementary Material to explain this difference. This model suggests that the existence of bulk states near the A-phase nodes modifies the critical velocity, compared to an isotropic superfluid, by a factor of $\sin(\theta_{\textrm{max}})\approx 0.57-0.85$ depending on pressure. Since the A-phase gap is proportional to $\sin(\theta)$, the low energy states near the node preferentially fill up first. The value $\theta_{\textrm{max}}$ is introduced as an effective quantity that sets the minimum energy of available bulk states for surface states to scatter into. The combination of this correction factor, and the corner flow enhancement, accounts for the relatively low critical velocity measured in our experiment.

In conclusion, we have carried out measurements of the force-velocity curves for oscillatory flow in $^3$He-A in channels with thicknesses of 1800 and 750 nm. We find dissipation onsets at a critical velocity that has the same temperature scaling as the Ginzburg-Landau gap. This critical velocity is best explained by the pumping of surface bound states in our devices, an effect that has not previously been shown in $^3$He-A. Our experiment studies channel sizes that are still large compared to the gap suppressed region where bound states are localized. In this regime, our measurement of the critical velocity does not show any notable dependence on the channel height or pressure.

In future experiments, we are interested in investigating the confinement limit where the channel thickness is comparable to -- or even smaller than -- the coherence length, a regime that is beginning to be experimentally accessible, such as in NMR devices of 192 nm \cite{heikkinen2021fragility} andHelmholtz resonators as small as 25 nm \cite{varga2022surface}. Theoretical work suggests that the A-phase is favored over the planar phase even in this highly confined limit when strong coupling is considered \cite{yapa2022triangular,sun2023superfluid}. In this highly confined regime, the gap suppression should extend across the entire channel, which we expect to have consequences for the critical velocity. Using the pressure and temperature dependence of the coherence length, $\xi(P,T)$, gives us an in situ knob to change the ratio $D/\xi$. Studying the Helmholtz mode force-velocity curves as a function of this ratio is thus a platform for probing the properties of bound states in both $^3$He-A and $^3$He-B, which are predicted to be exotic Weyl \cite{wu2023weyl,shevtsov2016electron,volovik2017chiral,silaev2012topological} and Majorana quasiparticles \cite{tsutsumi2012edge,wu2013majorana}.

\begin{acknowledgments}
The authors acknowledge that the land on which this work was performed is in Treaty Six Territory, the traditional territories of many First Nations, Métis, and Inuit in Alberta. They acknowledge fruitful discussions with C. Sun and F. Marsiglio, as well as support from the University of Alberta and the Natural Sciences and Engineering Research Council, Canada (Grant Nos.~RGPIN-2022-03078,  CREATE-495446-17, RGPIN-2021-02534, and DGECR2021-00043).
\end{acknowledgments}


%

\end{document}


\preprint{APS/123-QED}

\title{Supplementary Material}

\author{Alexander J. Shook}
\email{ashook@ualberta.ca}
\author{Emil Varga}%
\author{Igor Boettcher}
\author{John P. Davis}%
\email{jdavis@ualberta.ca}
\affiliation{%
Department of Physics, University of Alberta, Edmonton, Alberta, Canada T6G 2E9
}%


\maketitle

\section{Capacitance Bridge}

Our measurement makes use of a General Radio 1615-A, which has been described in previous publications \cite{rojas2015superfluid,varga2020observation,shook2020stabilized,varga2021electromechanical}. When the capacitance of the Helmholtz resonator satisfies the balance condition of the bridge, the current out of the detector port is zero. Driving the Helmholtz resonator near resonance creates a small periodic fluctuation
\begin{equation}
    C(t) = C_0 + \delta C(t).
\end{equation}
The current out of the detector is proportional to the time derivative of this capacitance fluctuation
\begin{equation}
    I_{\textrm{DET}} = \delta \dot{C}(V_{DC}+V_{AC}).
\end{equation}
In the linear drive regime, $\delta C$ is proportional to the electrostatic  drive force,
\begin{equation}
    \delta C(t) \propto F_E(t) = \frac{C_0}{2D}(V_{DC}+V_{AC}(t))^2.
\end{equation}
This means that the detector current is $I_{\textrm{DET}} \propto (V_{DC}+V_{AC})^3$, which produces frequency mixing. If the drive frequency is $\omega$ then $1\omega$, $2\omega$, and $3\omega$ terms are produced. However, only the $1\omega$ signal is studied as the detector current is demodulated at this frequency. Higher frequency terms can therefore be omitted without loss of accuracy 
\begin{equation}
    I_{\textrm{DET}} = \delta \dot{C}V_{DC},
    \label{eq:lin_I}
\end{equation}
\begin{equation}
    F_E = \frac{C_0}{2D} V_{DC} V_{AC}.
\end{equation}

\section{Mass Current Model}

The goal of this analysis is to show how the detector current responds to mass flow in the channels. The mass continuity equation requires that 
\begin{equation}
    2\oint \vec{j}_s \cdot d\vec{a} = -\frac{dM}{dt},
\end{equation}
where $j_s$ is the mass flux passing through the channel cross-sectional area, $a$, and $M$ is the fluid mass in the basin. The factor of $2$ accounts for the fact that there are two channels. Since the mass flux can change across the width of the slab, we introduce the notation
\begin{equation}
    \langle j_s \rangle = \frac{1}{D}\int_0^D j_s(z) dz.
\end{equation}
Equation \ref{eq:lin_I} can then be re-written as
\begin{equation}
    I_{\textrm{DET}} = V_{DC} \frac{dC}{dM} \frac{dM}{dt} = -\frac{V_{DC}}{2} \frac{dC}{dM} \langle j_s \rangle a.
\end{equation}
The basin mass is a function of both the volume
\begin{equation}
    V_B(t)=A(D-2\delta z(t)),
\end{equation}
and the density
\begin{equation}
    \rho(t) = \rho_0 + \delta \rho(t),
\end{equation}
where the small fluctuation is due to fluid compression. A small variation in mass therefore can be written as
\begin{equation}
    \delta M = AD \delta \rho - 2\rho_0 A \delta z.
    \label{eq:dM}
\end{equation}
The capacitance couples to the change in mass either through the displacement of the capacitor plates or modification of the helium permittivity due to compression
\begin{equation}
    C = \frac{\epsilon(\rho) A}{D-2\delta z} \approx \frac{\epsilon(\rho) A}{D} \left(1 + \frac{2\delta z}{D} \right),
    \label{eq:Cap}
\end{equation}
so the chain rule gives
\begin{equation}
    \frac{dC}{dM} = \frac{\delta\rho}{\delta M} \frac{d\epsilon}{d\rho} \frac{dC}{d\epsilon} + \frac{\delta z}{\delta M} \frac{dC}{dz}.
\end{equation}
Computing the capacitance derivatives from equation \ref{eq:Cap}, and using the Clausius-Mossoti relation
\begin{equation}
    \frac{d\epsilon}{d\rho} = \frac{\epsilon-1}{\rho},
\end{equation}
we arrive at
\begin{equation}
    \frac{dC}{dM} = \frac{C_0}{\rho} \left(\frac{\epsilon-1}{\epsilon}\right) \frac{\delta \rho}{\delta M} + \frac{2C_0}{D} \frac{\delta z}{\delta M}.
    \label{eq:Cap_response}
\end{equation}
Equation \ref{eq:dM} can be used to eliminate one variable
\begin{equation}
    \frac{\delta z}{\delta M} = -\frac{1}{2\rho_0A}\left(1- AD\frac{\delta \rho}{\delta M} \right).
    \label{eq:plate_fluid_coupling}
\end{equation}
Then a second equation is required, namely the balance of forces on the plate
\begin{equation}
    F_E = k_p\delta z + A\delta P.
    \label{eq:plate_EOM}
\end{equation}
Here $k_p$ is the stiffness of the quartz, and $\delta P$ is the pressure differential between the basin and the external helium bath. This can be related to the change in density, $\delta \rho = \rho \kappa_T \delta P$, where $\kappa_T = (\partial \rho / \partial P)_T$ is the isothermal compressibility. We neglect the plate inertia in equation \ref{eq:plate_EOM} because the applied drive frequency is much smaller than the resonance frequency of the quartz plate. Combining equations \ref{eq:dM} and \ref{eq:plate_EOM} to eliminate the $\delta z$ dependence, then differentiating gives
\begin{equation}
    \frac{\delta \rho}{\delta M} = \frac{1}{AD} \frac{2\Sigma}{1+2\Sigma},
\end{equation}
where we define the dimensionless ratio $\Sigma = \kappa_T k_p D/4A$. Plugging this result back into equation \ref{eq:Cap_response}
\begin{equation}
    \frac{dC}{dM} = \frac{C_0}{\rho AD} \left(\frac{\epsilon-1}{\epsilon}\right) \frac{2\Sigma}{1+2\Sigma} - \frac{C_0}{\rho_0 A D} \left(\frac{1}{1+2\Sigma} \right).
\end{equation}
Since $\Sigma$ and $\epsilon-1$ are both small quantities, the first term (electrostriction) can be dropped. In this case, we get an expression for the detector current
\begin{equation}
    I_{\textrm{DET}} = \frac{C_0 V_{DC}}{2\rho_0 A D} \left(\frac{1}{1+2\Sigma} \right) \langle j_s \rangle a.
\end{equation}
Finally, the detector current is related to a voltage by the convention factor, $R_{\textrm{trans}}$, set by the transimpedance amplifier
\begin{equation}
    \langle j_s \rangle  = \frac{(1+2\Sigma)}{2C_0R_{\textrm{trans}}} \left( \frac{\rho_0 A D}{a} \right) \frac{V_{\textrm{DET}}}{V_{\textrm{DC}}}.
\end{equation}

\section{Helmholtz Resonator Frequency}

The Helmholtz resonator frequency can be calculated explicitly when the resonator is in the linear flow regime $j_s = \rho_s v_s$. In this case the mass continuity equation yields
\begin{equation}
    2\rho_s a v_s = -\delta \dot{M} = 2\rho A \delta \dot{z} - \rho AD \kappa_T \delta \dot{P}.
\end{equation}
By taking the derivative and substituting in the second derivative of equation \ref{eq:plate_EOM} we can show
\begin{equation}
    \frac{a}{A} \left(\frac{\rho_s}{\rho}\right) \frac{\dot{v}_s}{D} = \frac{\ddot{F}_E}{k_pD} - (1+\Sigma) \frac{A}{k_pD} \delta \ddot{P}.
\end{equation}
The final equation required to derive the fourth sound resonance is the superfluid acceleration equation
\begin{equation}
    \dot{v}_s = \nabla \mu = \frac{\nabla P}{\rho} \approx \frac{\delta P}{\rho \ell_c},
\end{equation}
where $\ell_c$ is the length of the channel. The pressure equation of motion then is 
\begin{equation}
    \delta \ddot{P} + \omega_0^2 \delta P = \left(\frac{1}{1+\Sigma}\right) \frac{\ddot{F}_E}{k_pD},
\end{equation}
where the resonant frequency is 
\begin{equation}
    \omega_0^2(T) = \left(\frac{a}{\rho A^2 \ell_c}\right)\left(\frac{k_p/2}{1+\Sigma} \right) \left( \frac{\rho_s(T)}{\rho} \right).
\end{equation}
The maximum frequency of the Helmholtz resonator occurs when $\rho_s/\rho = 1$
\begin{equation}
    \omega_0^2(0) = \left(\frac{a}{\rho A^2 \ell_c}\right)\left(\frac{k_p/2}{1+\Sigma} \right).
\end{equation}
Because the superfluid suppression is negligibly small for the 1800 nm device, it can be accurately equated with the bulk temperature dependence \cite{ichikawa1987healing,freeman1990size,zhelev2017ab}. Accounting for Fermi liquid corrections, the temperature as a function of resonator frequency is
\begin{equation}
    \frac{T}{T_{c,1800}} = Y_0^{-1}\left(\frac{1-(\omega_0(T)/\omega_0(0))^2}{1+\frac{1}{3}F_1^s(\omega_0(T)/\omega_0(0))^2}\right),
\end{equation}
where $Y_0^{-1}$ is the inverted Yoshida function and $F_1^s$ is a Fermi liquid parameter \cite{Vollhardt2013}. We define $T_{c,1800}$ as the critical temperature of the 1800 nm device.

\section{Flow Enhancement}

The geometry of the Helmholtz resonator is such that the corners between the channel and basin are sharp relative to the size of the coherence length. When fluid flows into the basin from the channel or vice versa, there must be flow around the corners. This is significant because the local velocity around a corner will be higher than the mean velocity in the channel. To estimate the magnitude and spatial extent of this local flow enhancement we will idealize the flow as being incompressible. For incompressible potential flow the equation of continuity states that
\begin{equation}
    \vec{\nabla} \cdot \vec{v}_s = \nabla^2 \phi = 0,
\end{equation}
where $\phi$ is the velocity potential. This equation must be supplemented by boundary conditions which require the velocity normal to the solid surfaces to be zero. For a corner with an outer angle $\alpha_c$ there is an exact solution for the velocity potential near the corner in two dimensions \cite{landau2013fluid}
\begin{equation}
    \phi_{\textrm{local}}(r,\theta) = Ar^{n}\cos(n\theta),
\end{equation}
where $n=\pi/\alpha_c$, and the coordinate system is centered on the corner in question. The integration constant $A$ has units of m$^{2-n}$/s, so it can be separated into a velocity scale, $v_0$, and a length scale, $L$, such that $A=v_0L^{1-n}$. Taking the norm of the gradient gives the magnitude of the local velocity as a function of $r$
\begin{equation}
    |v_{\textrm{local}}| = v_0\left(\frac{L}{r} \right)^{1-n}.
\end{equation}
The angle $\alpha_c$ can be obtained from the intersection of the circle tangent line with the channel
\begin{equation}
    \alpha_c = \pi + \cos^{-1}\left(\frac{w}{2R}\right) \approx 4.482 \textrm{ [rad]},
\end{equation}
where $R=3.5$ mm is the basin radius, and $w= 1.6$ mm is the channel width, such that $1-n \approx 0.299$.

\begin{figure}
    \centering
    \includegraphics[width=\linewidth]{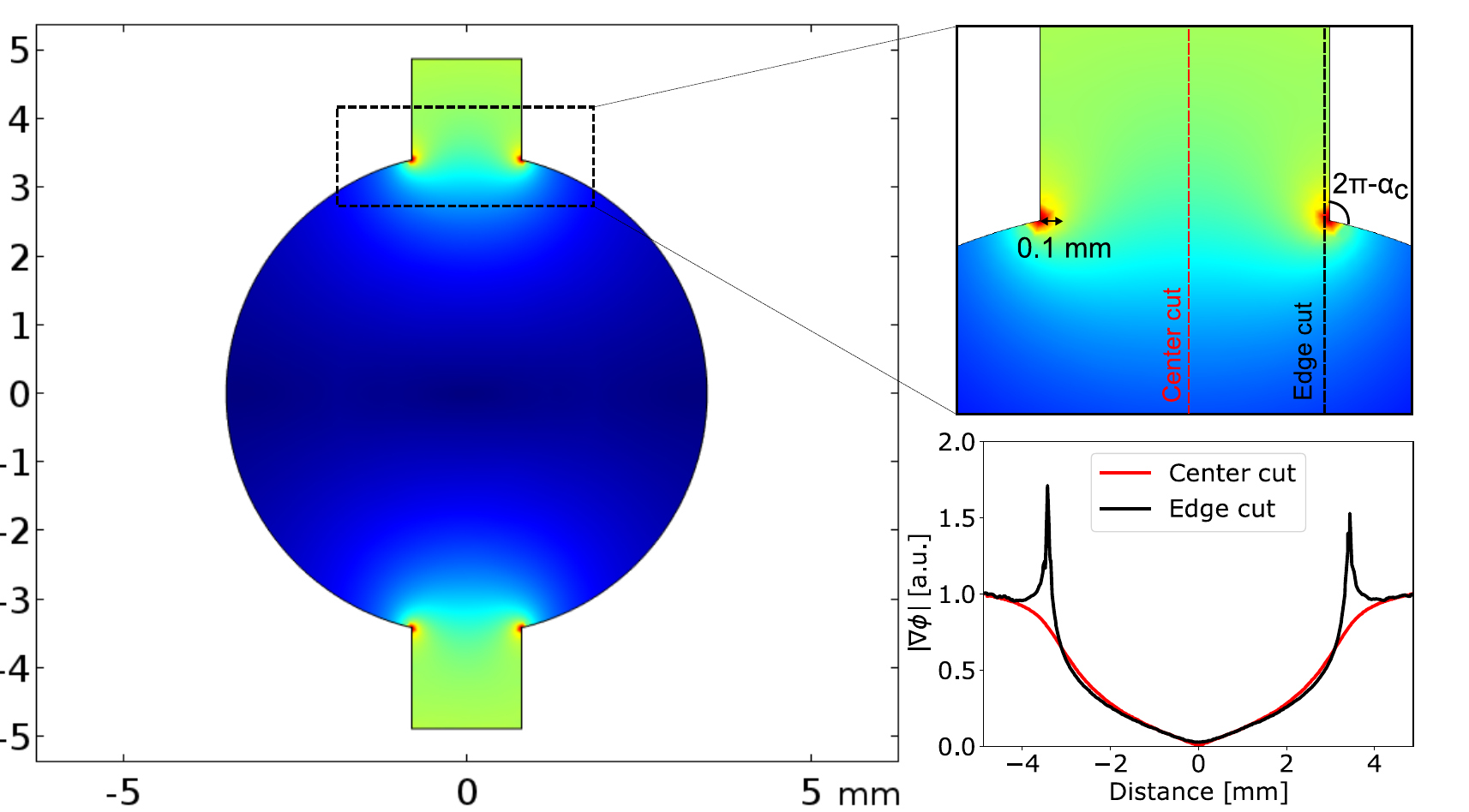}
    \caption{Finite element simulation solving the velocity potential equation $\nabla^2\phi=0$ for our device geometry. The color scale represents the magnitude of the velocity $v = |\vec{\nabla}\phi|$. Dirichlet boundary conditions are imposed at the openings of the channels to create a finite phase difference between the basin and channel opening. Flow enhancement occurs at the corners due to the wall boundary conditions, which defines a length $L\sim 0.1$ mm over which the velocity decays to the mean channel velocity, $v_0$, which is set by the Dirichlet boundary conditions.}
    \label{fig:COMSOL}
\end{figure}

The constants $v_0$ and $L$ must be specified by the boundary conditions of the problem. To this end we have used the finite element solver COMSOL Multiphysics to solve Laplace's equation for the boundary conditions of our devices. We use this simulation to estimate the length over which the enhanced flow recovers the mean channel velocity, $v_0$, to be $L \sim 100$ $\mu$m. The flow is simulated by using Dirichlet boundary conditions at the openings of the channels to set the initial value of the velocity potential. Within the potential flow model, this is equivalent to imposing a pressure difference between the channel opening and basin. It should be noted that while the choice of initial velocity potential modifies the velocity scale, $v_0$, it does not change $L$, which is only a function of the wall geometry. The ratio $|v_{\textrm{local}}|/v_0$ is therefore purely geometric, such that no material parameters appear in the simulation.

The divergence of the velocity as $r\to 0$ is cutoff by a minimum length scale over which order parameter gradients can exist, which is taken to be the zero temperature coherence length. This defines a maximum velocity 
\begin{equation}
    \frac{|v_{\textrm{max}}|}{v_0} \sim \left(\frac{100 \textrm{ }\mu\textrm{m}}{\xi_0(P)}\right)^{0.299}.
\end{equation}
The size of the coherence length and flow enhancements are summarized in Table 1 for the pressures relevant to our experiment. It is worth noting that due to the magnitude of the exponent, the flow enhancement is not too sensitive to small changes in the coherence length or small errors in the estimate of $L$. Increasing $L/\xi_0$ by a factor of $10$ only results in the flow enhancement changing by a factor of $2$. Therefore, we expect the flow enhancement estimate to be reasonable provided that $L$ is the correct order of magnitude. Since $L$ cannot be larger than the channel width, which is on the order of millimeters, the flow enhancement cannot be off by more than a factor of 2. 

\begin{table}[b]
\begin{tabular}{|c|c|c|}
\hline
Pressure {[}bar{]} & $\xi_0$ {[}nm{]} & $|v_{\textrm{max}}|/v_0$ \\ \hline
2.87                  & 51               &  9.62                    \\ \hline
22.45                 & 20               & 12.83                    \\ \hline
27.94                 & 17               & 13.31                     \\ \hline
\end{tabular}
\label{tab:flow_enhancement}
\caption{Zero temperature coherence length and flow enhancement at different pressures.}
\end{table}
Since the flow velocity at the corners is an order of magnitude higher than the rest of the channel, pair-breaking will first onset near the corners when sweeping the drive power. Although the region of flow enhancement is relatively small, a theoretical calculation performed by C. J. Lambert \cite{lambert1992theory} suggests that the power dissipated due to surface pair breaking scales non-linearly, meaning that the corner flow may dominate the total dissipated power. If this is the case then the true critical velocity is $|v_{\textrm{max}}|$ rather than $v_0$. At present, we cannot control the degree of flow enhancement, so this remains speculation. A future experiment however could vary the flow enhancement by studying devices with increasingly rounded corners.

\section{A-B Phase Boundary}

For the measurements conducted at 22 and 28 bar, the fluid inside the Helmholtz resonator is A-phase, however the surrounding bath is B-phase. It has been pointed out that the resulting A-B domain wall may oscillate in response to the fluid flow generating dissipation \cite{buchanan1986velocity}. The power dissipated by a moving A-B interface is 
\begin{equation}
    \dot{Q}_{\textrm{DW}} = L S u
\end{equation}
where $L$ is the latent heat per unit volume, $S$ is the domain wall surface area, and $u$ is the velocity of the domain wall \cite{bartkowiak2002unique}. If the domain wall oscillates in phase with the velocity field such that $u\propto v_s$, then we expect that this would not result in a critical velocity, but would simply broaden the Helmholtz resonance. Nevertheless, we chose to test this proposal by performing additional measurements at 28 bar at temperatures above $T_{\textrm{AB}}=2.11$ mK, such that the bulk fluid is A-phase. If the onset of dissipation is attributable to the motion of the domain wall, we would then predict the critical behavior and non-linear distortion of the line shape to disappear. Contrary to this prediction we find the same inflection point in the the force-velocity curve following the same $\sqrt{1-T/T_c}$ temperature dependence even in the absence of an A-B domain wall. Figure 2 shows the force-velocity curve measured at 28 bar and $T=2.37$ mK.
\begin{figure}
    \centering
    \includegraphics[width=\linewidth]{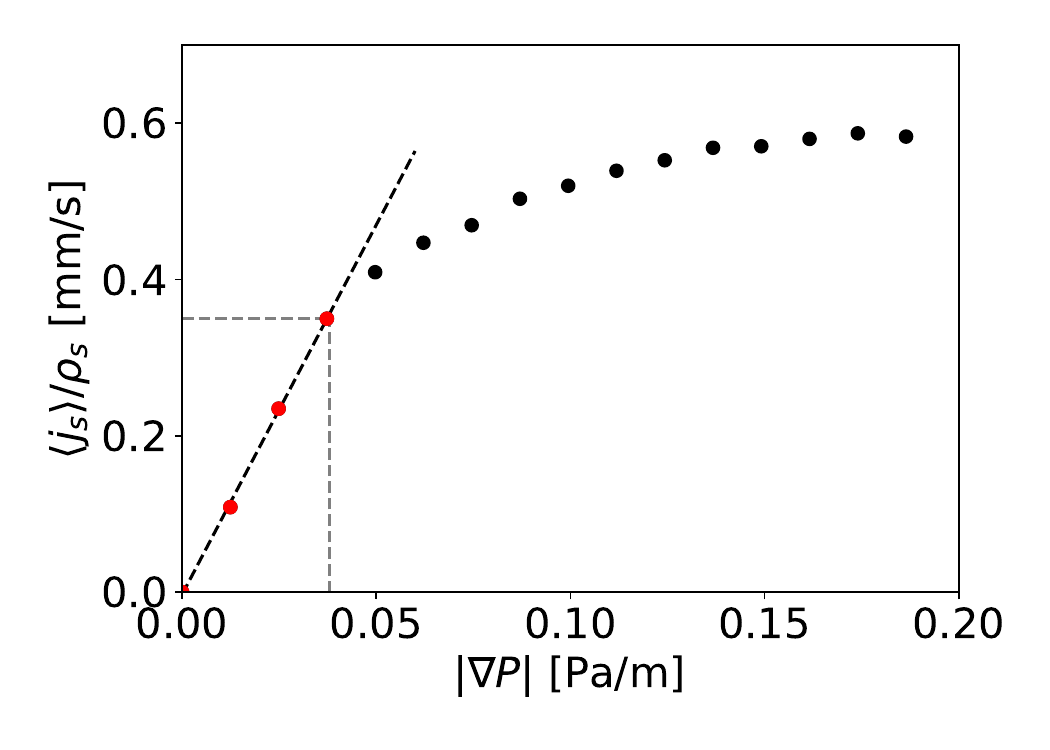}
    \caption{Velocity force curve for the 750 nm Helmholtz resonator measured at 28 bar at 2.37 mK. The temperature of the experiment is about the bulk $T_{AB}$ line such that the superfluid is A-phase everywhere in the experimental cell. The red points indicate the linear regime, which has been fit to a dashed line. The veritcal and horizontal dashed lines indicate the last measured point of the linear regime.}
    \label{fig:F-v_curve}
\end{figure}

\section{Textural Gradients}

As outlined in the main text, an alternative interpretation for the critical velocities observed in the A-phase is that they could be due to a secondary texture becoming energetically favorable past a particular velocity threshold. This type of critical velocity can be estimated by comparing terms in the Ginzburg-Landau free energy. For the A-phase in the weak coupling limit, the free energy density due to gradients in the order parameters can be separated into three parts \cite{Vollhardt2013}
\begin{equation}
    f_{\textrm{bend}} = f_{\phi} + f_{\phi,\ell} + f_{\ell} + f_{\ell, d},
\end{equation}
where $f_{\phi} = \rho_s v_s^2$ includes only phase gradient (i.e., velocity) terms, $f_{\ell}$ includes only orbital textural gradients, $f_{\phi,\ell}$ includes terms coupling phase and textural gradients, and $f_{\ell d}$ includes coupling to gradients in the spin anisotropy vector, $\hat{d}$,
\begin{equation}
    f_{\phi,\ell} = - \frac{\rho_s}{2}(\hat{\ell}\cdot\vec{v}_s)^2 + \frac{\hbar\rho_s}{2m} \left[\frac{1}{2}\vec{v}_s\cdot(\vec{\nabla}\times\hat{\ell})-(\hat{\ell}\cdot\vec{v}_s)\hat{\ell}\cdot(\vec{\nabla}\times\vec{v}_s) \right],
\end{equation}
\begin{equation}
    f_{\ell} = \frac{\rho_s}{4}\left(\frac{\hbar}{2m}\right)^2\left[(\vec{\nabla}\cdot\hat{\ell})^2+[\hat{\ell}\cdot(\vec{\nabla}\times\hat{\ell})]^2+3[(\hat{\ell}\cdot\vec{\nabla})\hat{\ell}]^2\right],
\end{equation}
\begin{equation}
    f_{\ell, d} = 2\rho_s \sum_{\alpha} \left[ 2[(\hat{\ell}\times\vec{\nabla})\hat{d}_{\alpha}]^2 + [(\hat{\ell}\cdot\vec{\nabla})\hat{d}_{\alpha}]^2 \right].
\end{equation}
The boundary conditions require that $\hat{\ell}$ be parallel to the surface normal $\hat{n}$ at the wall. For a slab geometry confined between two infinite planes with zero flow, the lowest energy texture is simply one with $\pm \hat{\ell} \parallel \hat{n}$ everywhere. For a system with finite superflow the negative terms in $f_{\phi,\ell}$ reduce the free energy when $\hat{\ell} \parallel \vec{v}_s$. For a sufficiently high velocity, the $\hat{\ell}$ field will bend towards $\vec{v}_s$ in the middle of the slab while still satisfying the boundary conditions at the walls. Textural gradients increase the positive free energy terms, which results in a kind of textural rigidity opposing the flow alignment effect. When the negative flow alignment terms become greater than the positive textural gradient terms, a phase transition occurs at a critical velocity which is analogous to the Fréedericksz transition in liquid crystals \cite{freedericksz1933forces}. de Gennes and Rainer \cite{de1974alignment} calculated this critical velocity to be
\begin{equation}
    v_{\textrm{Fr}} = \sqrt{\frac{3}{4}} \frac{\pi\hbar}{2mD},
\end{equation}
assuming no gradients in $\hat{d}$, which is 38 mm/s for the 750 nm device and 16 mm/s for the 1800 nm device. This is an order of magnitude higher than the mean channel velocity but may be comparable to the velocity very close to the corners. The critical velocity $v_{\textrm{Fr}}$ assumes that gradients only exist in the highly confined dimension, such that the ``tilting" of the texture occurs across the entire slab. If the corner flow enhancement is responsible for a phase transition in the texture, the change must instead be localized to within a few coherence lengths of the corner where the velocity attains its maximum. At this distance, one must consider not only the effects of the top and bottom walls confining the slab but the side walls as well. The texture in the region near the side walls is potentially complicated. Despite this, it is safe to say that the close proximity to the side wall means that such a critical velocity must be significantly higher than $v_{\textrm{Fr}}$. For the Fréedericksz-like transition the gradient is over a characteristic length scale of $D\approx 1$ $\mu$m rather than a few coherence lengths ($\approx 10^{-2}$ $\mu$m) as would be the case for the localized corner transition. Furthermore, for the same reasons outlined in section IV, it is unlikely that such a localized effect could significantly modify the total mass current.

Finally, we consider the possibility of textural domain walls in our devices. Because the boundary conditions require that $\hat{\ell}$ be either parallel or anti-parallel to the surface normal, it is possible to have a boundary where $+\hat{\ell} \parallel \hat{n}$ on one side and $-\hat{\ell} \parallel \hat{n}$ on the other \cite{kasai2018chiral,walmsley2004intrinsic}. Such defects can be created during the phase transition from a higher symmetry phase (i.e., the normal fluid or B-phase) to a lower symmetry phase (i.e., the A-phase) via the Kibble-Zurek mechanism \cite{zurek1985cosmological}. The dissipative force felt by the fluid due to the defects is the Magus force, which is proportional to $\vec{v}_s \times \hat{\ell}$ \cite{ikegami2013chiral}. In practice, the dissipation will depend on a spatial average of the texture which will set in stochastically every time the A-phase is nucleated. As is argued in section V, a dissipative mechanism that is linear in the flow velocity will merely broaden the resonance. To produce a critical velocity the domain wall structure must change above some threshold velocity. If this is the case, we would expect to see hysteric behavior which changes both over multiple power sweeps, and over multiple nucleations of the A-phase. Our characterization of the critical velocity however is reproducible over multiple cooldowns and does not show hysteretic behavior after repeated power sweeps.


\section{Vortex Dissipation}

The high degree of confinement used in our experiment makes the formation of vortices involving disgyration of the orbital angular momentum vector $\hat{\ell}$ unlikely. Hard-core vortices due to phase winding are still possible however. Remnant vortices produced during a phase transition may become pinned inside the channel or basin. Pinned hard-core vortices contribute a dissipation that is linear in velocity due to mutual friction, up until the de-pining velocity is exceeded, at which point the vortices are annihilated.

Additional vortices can be nucleated when the velocity exceeds the intrinsic instability velocity $v_{cb}$. In the Ginzburg Landau regime this critical velocity can be shown to be \cite{ruutu1997intrinsic}
\begin{equation}
    v_{cb} = 1.61(1+F_1^s/3) \frac{k_BT_c}{p_F}\sqrt{1-T/T_c}.
\end{equation}
The zero temperature value is $v_{cb}(0)=$70-312 mm/s for pressures ranging from 0 to 28 bar. The zero temperature critical velocity extracted from the fit to our data is 2.65 mm/s, and appears to be independent of pressure for pressures over a wide pressure range of 2.87 to 27.94 bar. Since the average channel velocity is low, vortex formation must happen a points of high localized velocity due either to sharp angles in the geometry (i.e. the channel corners), or surface roughness. We have characterized the roughness of our devices using an Alpha Step IQ surface topography profiler, and found that the roughness is at most $\pm 9$ nm. Since this is small compared to the core size of $^3$He vortices, we conclude the flow enhancement due to roughness is insignificant to the corner flow enhancement. As discussed in section IV, this enhancement factor is $\sim 10-20$ but cannot be much larger.

Deep in the non-linear drive regime, the momentum density reaches approximately three times the critical value before starting to saturate, so the maximum corner velocity at zero temperature can be predicted to be $20 \times 3 \times 2.65$ mm/s $\approx 160$ mm/s. Based on these numbers we conclude that while vortices could be nucleated at the corners deep in the non-linear regime, vortex formation cannot account for the onset of the non-linear regime. Not only is the flow enhancement too small, but the large predicted pressure dependence is inconsistent with our measurements.

Another vortex dissipation process thought to occur due to superfluid flow through small orifices is a vortex mill process. Theoretical work first preformed in $^4$He by Schwarz \cite{schwarz1990phase} suggests that for superfluid flow through an orifice, continuous dissipation can be generated by a process where one end of the vortex remains pinned (presumably due to it being in a region of low flow velocity), while the other end exits the orifice, develops waves, crosses the flow lines of the channel, and then reconnects to a wall at the opposite side of the orifice. This process is thought to be connected to turbulence experiments in $^3$He-B using an oscillating grid \cite{bradley2008grid}.


A few general remarks should first be made about the geometry and flow field of the experiment. The velocity required to depin a vortex line from a protrusion of size $b$ is expected to be \cite{barenghi2001quantized}
\begin{equation}
    v_v = \frac{\kappa}{2\pi R} \ln\left(\frac{b}{\xi_0}\right).
\end{equation}
Here $\kappa$ is the quanta of circulation, $R$ is the length of the vortex line, and the vortex core size is estimated to be $\xi_0$. Given that any existing protrusions cannot be larger than the confined dimension $D \approx 10-100 \xi_0$, the logarithmic factor is, at most, of order unity. Since the vortex mill scenario requires the mobile end of the vortex line to cross the entire channel width, the vortex must span at least 1.6 mm during the process. This implies a depinning velocity $v_v \approx 6.6$ $\mu$m/s, which is much smaller than the characteristic experimental velocities of $\sim 1$ mm/s. The static end of the vortex must therefore exist far from the channel where the flow field is sufficiently low. 

\begin{figure}
    \centering
    \includegraphics[width=0.8\linewidth]{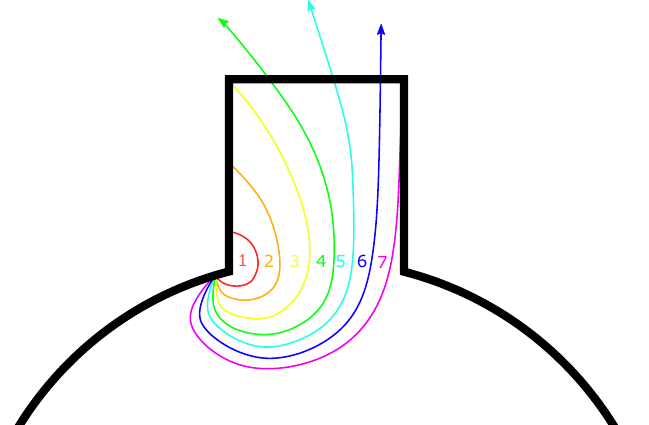}
    \caption{An illustration of a hypothetical vortex line motion that might produce a vortex mill similar to that described by Schwarz. The vortex has one static end pinned inside the basin, and a mobile end which at some point in the process exits the channel.}
    \label{fig:Schwarz_Cartoon}
\end{figure}

There are two available regions of low flow velocity; namely inside the basin and outside of the Helmholtz resonator. Since the static end of the vortex cannot exist inside the channel, the vortex line needs to cross both the length and width of the channel. All of this must occur without the vortex line intersecting the top and bottom walls that constrain the confined dimension. This means essentially all the motion occurs in a plane parallel to the wall. Figure \ref{fig:Schwarz_Cartoon} depicts the type of trajectory which would need to occur to realize this vortex mill process. Such a trajectory however is not physically plausible. The equation of motion for a vortex line expressed as a parameterized curve $\vec{s}(\xi,t)$, is given by Schwarz's equation. Assuming the normal fluid is clamped, and neglecting the vortex self-induced velocity, Schwarz's equation is \cite{walmsley2004intrinsic}
\begin{equation}
    \frac{d\vec{s}}{dt} = (1 - \alpha') \vec{v}_s + \alpha \vec{s}\mkern2mu\vphantom{s}' \times \vec{v}_s.
\end{equation}
Here, $\vec{s}\mkern2mu\vphantom{s}'$ is the derivative of the curve with respect to the arclength $\xi$. Assuming flow in the channel is mostly uniform in the $x$ direction, this gives
\begin{equation}
    \frac{d\vec{s}}{dt} = (1 - \alpha') v_{s,x} \hat{x} + \alpha s_z' v_{s,x} \hat{y} - \alpha s_y' v_{s,x} \hat{z}.
\end{equation}

The shape of the curve is of course not known. What is of significance is that the velocity in the two non-flow directions are of comparable magnitude. It is not clear how the mobile end of the vortex line can cross a distance of millimeters in say the $x$ and $y$ directions, while the entire length of the line moves less than a micron in the $z$ direction in response to the driven flow.

Based on the above arguments, the vortex mill process seems physically plausible for orifices with lengths comparable to their diameter, such as the grid studied by the Lancaster group ($40$ $\mu$m square holes with a depth of $11$ $\mu$m) \cite{bradley2008grid}, but not for the long, wide, and flat channels used in our devices.


The final, and perhaps most important, argument against vortex dissipation being responsible for the critical velocities observed in our experiments is that it fails to account for the difference we observe between the A and B phases. In the B-phase, the linear drive regime exists up to velocities about three times larger than the A-phase. When B-phase exists in the Helmholtz resonators, the non-linear behavior is of a qualitatively different kind. The resonance line shape, instead of flattening, develops a bistability. Figure 5 in the main text shows how the resonance peak height scales with drive voltage for temperatures near the A-B critical temperature. If the A-phase critical velocity was due to pure phase winding vortex dissipation, we would expect it to also exist in the B-phase. Since the B-phase does not exhibit this behavior, we conclude that the critical velocity is not due to vortex dissipation.




\section{Pumping Excitations}

In the BCS theory of superfluidity, which we apply here to $^3$He, the elementary excitations are Bogoliubov quasi-particles (BQP) \cite{Vollhardt2013}. For a superfluid flowing at velocity $v_s$ the dispersion relation for BQP is 
\begin{equation}
    E_{\pm}(p,\theta,z,v_s) = \vec{p} \cdot \vec{v}_s \pm \sqrt{\xi_{\vec{p}}^2+\Delta^2(z,\theta)}.
\end{equation}
Here, the $\pm$ signifies the two branchs of the dispersion relation, $\xi_{\vec{p}} = (p^2-p_F^2)/(2m^*)$ is the energy of a Fermi liquid excitation, with $p_F$ being the Fermi momentum, $m^*$ the effective mass, and $\Delta$ is the superfluid gap, which we allow to vary with distance from a wall $z$. The A-phase gap also has an angular dependence $\Delta(z,\theta) = \Delta_A(z)|\sin(\theta)|$, where $\theta = \cos^{-1}(\hat{p} \cdot \hat{\ell})$.

For $v_s=0$ conservation of energy prevents excitations in the lower branch from scattering into the upper branch. For finite $v_s$, however, the Galilean boost term, $\vec{p} \cdot \vec{v}_s$, lifts the degeneracy in the $xy$-plane such that excitations with momentum parallel to $\vec{v}_s$ have higher energy than excitations with momentum anti-parallel to $\vec{v}_s$. The threshold velocity at which excitations from the lower branch can scatter into the upper branch is when min$(E_{+})$ $=$ max$(E_-)$. Assuming $v_s > 0$, the minimum of the upper branch occurs at $p=-p_F$, and the maximum of the lower branch is at $p=+p_F$. The critical velocity criterion therefore is 
\begin{align}
\nonumber
0 & = E_{+}(-p_F,\theta_1,z_1,v_c)-E_{-}(+p_F,\theta_2,z_2,v_c) \\ & = -2p_Fv_c + |\Delta(z_1,\theta_1)| + |\Delta(z_2,\theta_2)|.
\end{align}
When this criterion is met, a branch conversion process can occur where excitations with momentum $p=+p_F$ scatter into the $p=-p_F$ state. For oscillatory flow, this allows for a process by which low-energy states with a suppressed gap can be continuously pumped into states where the gap is larger.

For a hypothetical superfluid where the gap is isotropic and spatially invariant such that $\Delta(z,\theta) = \Delta_0 =$ const., the threshold is given by $v_c = \Delta_0/p_F$, which is the Landau result. For a superfluid that is isotropic, but goes to zero for all directions at a surface (as would be the case in the B-phase with diffuse scattering conditions) the threshold is $v_c^B = \Delta_B^{\textrm{bulk}}/2p_F$. Considering the case where $\Delta$ is both suppressed completely at the wall and anisotropic (appropriate to the A-phase with diffuse scattering conditions) there are now two variables that cause the gap to decrease to zero.

In principle, it might be possible for the bulk (i.e. $z \gg 0$), near the nodes ($\sin\theta \ll 1$) to scatter into bulk states where the gap is larger due to the oscillatory flow. This would require quasiparticle-quasiparticle scattering to conserve momentum since by assumption it occurs in the bulk far from a wall. Such a process, however, does not constitute a critical velocity, since it can happen for arbitrarily small velocities. 

For a surface state in the A-phase to escape into the bulk the criterion is 
\begin{align}
\nonumber
0 & = E_{+}(-p_F,\theta_1,\infty,v_c)-E_{-}(+p_F,\theta_2,0,v_c) \\ & = -2p_Fv_c + \Delta_A^{\textrm{bulk}} |\sin(\theta_1)|.
\end{align}
In the presence of a wall at $z=0$, we compare excitations of the lower branch $E_-$ at $z_2\approx 0$ where the gap closes, $\Delta(z_2,\theta_2)=0$ (these are the surface states), to excitations of the upper branch $E_+$ occur in the bulk such that $\Delta(z_1,\theta_1)=\Delta_A^{\rm bulk}(z_1)|\sin(\theta_1)|$. This implies that the critical velocity is $v_c^A = \Delta_A^{\textrm{bulk}} |\sin(\theta)|/2p_F$. The threshold should then be set by an effective filling level of the bulk states near the node. Supposing all of the states for $\theta < \theta_{\text{max}}$ are filled, then the critical velocity would be $v_c^A = \Delta_A^{\textrm{bulk}} |\sin(\theta_{\text{max}})|/2p_F$. The variable $\theta_{{\text{max}}}$ is an effective value that should be set by some combination of the thermal occupation and flow considerations. It is worth noting that, unlike the experiments of Castelijns et al.~\cite{castelijns1986landau} and Bradley et al.~\cite{bradley2016breaking}, which took place in the ballistic regime, our experiment is performed at relatively high temperatures due to the need to stabilize the A-phase. Our measurements are all in the hydrodynamic regime where thermal excitations are significant. As pointed out by Castelijns et al.~\cite{castelijns1986landau}, it is possible at finite temperatures that quasiparticles can play an intermediate role in allowing the bound states to escape into the bulk.

\begin{figure}
    \centering
    \includegraphics[width=\linewidth]{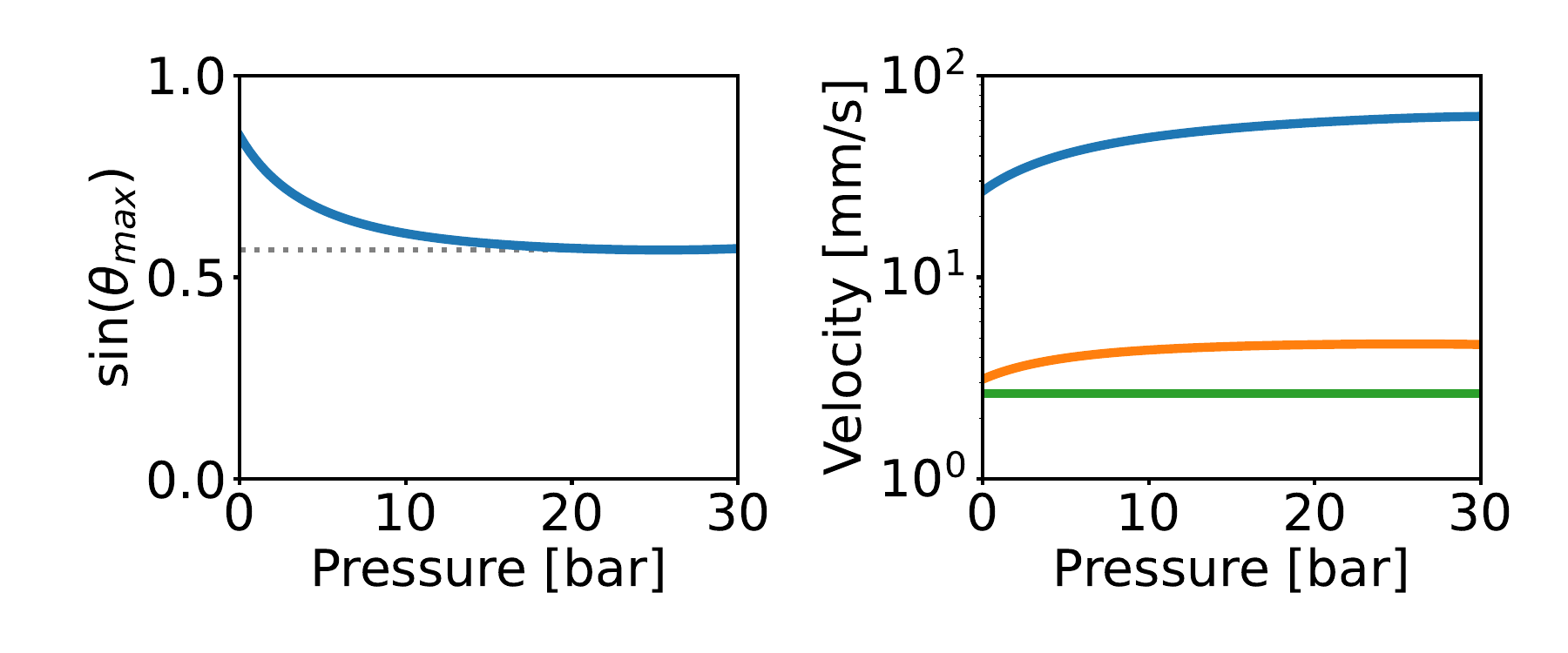}
    \caption{a) Plot of the estimated value of $\theta_{\textrm{max}}$ as a function of pressure. The dashed line shows the minimum value of 0.568 which applies at high pressures. b) Comparison of the expected pressure dependence of the critical velocity in three cases. One is the pressure dependence given by $v_c = \Delta(P)/p_F(P)$ (blue), the second is reduced by the flow enhancement factor $(L/\xi_0)^{0.299}$ (orange), and the third is the pressure independent fit parameter $v_0$ that we extract from our data (green).}
    \label{fig:Angle_analysis}
\end{figure}

Using our measured results we can estimate a value of $\theta_{{\text{max}}}$. For comparison to the data, we assume the maximum A-phase gap follows the Ginzburg-Landau temperature dependence,
\begin{equation}
    \Delta_A^{\textrm{bulk}}(T) = a_A k_BT_c \sqrt{1-T/T_c},
\end{equation}
where $a_A = 3.42$ \cite{Vollhardt2013}. Using the fit to the data, $v_c = (2.65 \textrm{ mm/s})\sqrt{1-T/T_c}$, and adjusting for the corner flow enhancement calculated in section IV, implies
\begin{equation}
    \sin(\theta_{{\text{max}}}(P)) \approx \frac{2p_F(P)}{a_Ak_BT_c(P)} (2.65 \textrm{ mm/s}) \left(\frac{100 \textrm{ }\mu\textrm{m}}{\xi_0(P)}\right)^{0.299}
\end{equation}
The pressure dependence of this function is shown in Fig.~4a. Our measurements suggest that the observed critical velocity is pressure-independent. The lack of pressure dependence is propagated into the estimate of $\sin(\theta_{{\text{max}}}(P))$, since it must then cancel the pressure dependence of the other parameters by assumption. Because our experiment does not allow for measurement of the flow enhancement, this estimation is only speculative at this stage. It could well be the case that the pressure dependence of the flow enhancement differs from the simple potential flow model used in Section IV. Likewise, the magnitude of $\sin(\theta_{{\text{max}}}(P))$ is sensitive to errors in the estimate of the flow enhancement. A more thorough investigation of pressure dependence will have to be left to future experiments in which the flow enhancement can be modified.


%